\documentclass[aps,prc,floats,floatfix,preprint,superscriptaddress,nofootinbib]{revtex4-1}  
\def\be{\begin{eqnarray}}   
\def\ee{\end{eqnarray}}

\newcommand{\affA}{
Graduate School of Pure and Applied Sciences, University of Tsukuba, 
Tsukuba 305-8571, Japan
}
\newcommand{\affB}{
Center for Computational Science, University of Tsukuba, 
Tsukuba 305-8571, Japan
}
\newcommand{\affC}{
Department of Medical and General Sciences, Nihon Institute of Medical Science,
1276 Shimogawara, Moroyama-machi, Iruma-gun, Saitama 350-0435, Japan
}
\newcommand{\affD}{
Max Planck Institute of Microstructure Physics, 06120 Halle, Germany
}

\usepackage{graphicx}
\usepackage{epstopdf}
\usepackage{color}

\begin{document}

\title{Nonlinear electronic excitations in crystalline solids
using meta-generalized gradient approximation and hybrid functional 
in time-dependent density functional theory
}

\author{Shunsuke A. Sato}\affiliation{\affA}
\author{Yasutaka Taniguchi}\affiliation{\affB}\affiliation{\affC}
\author{Yasushi Shinohara}\affiliation{\affD}
\author{Kazuhiro Yabana}\affiliation{\affA}\affiliation{\affB}

%========================================================================
\begin{abstract}
We develop numerical methods to calculate electron dynamics in
crystalline solids in real-time time-dependent density functional theory
employing exchange-correlation potentials which reproduce band gap 
energies of dielectrics; a meta generalized gradient 
approximation (meta-GGA) proposed by Tran and Blaha 
[Phys. Rev. Lett. 102, 226401 (2009)] (TBm-BJ) and 
a hybrid functional proposed by Heyd, Scuseria, 
and Ernzerhof [J. Chem. Phys. 118, 8207 (2003)] (HSE).
In time evolution calculations employing the TB-mBJ potential, 
we have found it necessary to adopt a predictor-corrector step for 
stable time-evolution. Since energy functional is not known for the TB-mBJ 
potential, we propose a method to evaluate electronic excitation energy 
without referring to the energy functional. Calculations using the HSE hybrid 
functional is computationally expensive due to the nonlocal Fock-like term.
We develop a computational method for the operation of the Fock-like term 
in Fourier space, for which we employ massively parallel computers 
equipped with graphic processing units.
To demonstrate significances of utilizing potentials providing
correct band gap energies, we compare electronic excitations induced 
by femtosecond laser pulses using the TB-mBJ, HSE, and a simple 
local density approximation (LDA).
At low laser intensities, electronic excitations are found to be
sensitive to the band gap energy: results using TB-mBJ and HSE are
close to each other, while the excitation of the LDA calculation 
is more intensive than the others.
At high laser intensities close to a damage threshold, we have found that
electronic excitation energies are similar among the three cases. 

\end{abstract}
\maketitle
%========================================================================
\section{Introduction}
%========================================================================

At current frontiers of optical sciences, interactions of
intense and ultra-short laser pulses with solids are attracting 
much interests \cite{br00,mo06,kr09}.  For example, ultra-short
laser pulses with intensities above a damage threshold
are used for non-thermal laser-processing  \cite{pe99, ba13}.
Ultra-short laser pulses irradiated on dielectrics close to 
the damage threshold have been found to show intriguing phenomena 
such as producing an ultrafast charge transfer \cite{sc13}, and
modifying the band gap in sub-femtosecond time scale \cite{ma14}.
These phenomena induced by ultra-short, intense laser pulses 
are expected to open up new technological applications.

For microscopic understanding of these phenomena,
it is quite significant to theoretically investigate nonlinear electron 
dynamics in the medium excited by the strong electric field of 
laser pulses. Real-time calculation based on the time-dependent density 
functional theory (TDDFT) \cite{ru84} is a useful theory for the investigation of 
nonlinear electron dynamics.
It has been applied to nonlinear laser-solid interactions such as optical 
breakdown \cite{ot05}, generation of coherent phonon \cite{sh10}, and 
calculation of nonlinear optical properties \cite{go13}, as well as linear responses 
for a weak field \cite{ba00}.

In order to apply the TDDFT to practical problems, we need to use
approximate exchange-correlation potentials. In most applications
of the TDDFT, adiabatic approximation, in which the exchange-correlation 
potential at each time is determined by the density (, current, 
and kinetic density  if necessary) at the same time, is assumed.  
In the following developments, we will consider three
exchange-correlation potentials: local density approximation (LDA),
meta generalized gradient approximation (meta-GGA), and hybrid 
functional, and use them in both ground state and time evolution
calculations in the adiabatic approximation.

Among the three approximations, the LDA is the simplest one.
We will use the energy functional of Ref. \cite{pe81}.
Because of the well-known band gap problem in the
static LDA calculations, optical gaps of semiconductors and insulators 
are also underestimated in the TDDFT using the adiabatic LDA. 
The underestimation of the optical gap causes serious 
problems when we compare calculated physical quantities with 
experimental results in nonlinear electronic excitations induced 
by intense laser pulses as well as in linear responses for a weak optical fields.

Recently, Tran and Blaha proposed a meta-GGA exchange 
potential which systematically improves the band gap energy of various
materials \cite{tr09}. We abbreviate the potential as TB-mBJ potential.
Calculated band gap energies using the TB-mBJ potential agree 
with experimental gap energies  in a quality similar to those by 
the {\it GW} method \cite{tr09}.

Hybrid functional is, at present, one of the most successful 
approximations in practical applications of static and time-dependent
density functional theory.
In solid state physics, a hybrid functional that 
is proposed by the Heyd, Scuseria, and Ernzerhof (HSE) has been widely 
applied \cite{he03} .
The HSE functional significantly improves fundamental gap comparing to one
in the LDA. It has been shown that also accurate optical absorption spectra
in several dielectrics and partial inclusion of excitonic excitation by
J. Paier {\it et al} \cite{pa08}.

In this work, we present numerical methods to achieve time-evolution 
calculations in the real-time TDDFT employing the TB-mBJ
and HSE potentials.
We have already reported results employing the TB-mBJ 
potential  in the real-time TDDFT for light-matter interactions in 
both linear and nonlinear regimes \cite{sa14,sa14-2,wa14}.
To carry out stable time-evolution calculations,
we find that a predictor-corrector step is indispensable
in the calculations using the TB-mBJ potential, while it is not necessary
in the calculations using the LDA and HSE potentials.
The TB-mBJ potential is given in Ref. \cite{tr09} without referring 
to the energy functional, and an explicit form of the energy functional 
providing the TB-mBJ potential is not known. 
We discuss how to evaluate electronic
excitation energies during the irradiation of a laser pulse without
referring to the  energy functional.
In order to calculate nonlinear electron dynamics using 
the HSE functional, we develop a Fourier-space method for the
Fock-like operations: the non-local Fock-like term in the 
HSE functional is calculated in Fourier space, making use of
an efficient Fast-Fourier-Transformation library on an accelerator 
type super-computer.
We apply the above methods for crystalline silicon under irradiation
of ultrashort laser pulses, and compare electronic excitations by 
the three potentials.

The paper is organized as follows. In Sec. II, we describe 
methods of electron dynamics calculations in crystalline solids
utilizing the TB-mBJ and HSE potentials. 
In Sec. III and IV, we present calculations for electron dynamics in
crystalline silicon, using LDA, TB-mBJ, and HSE potentials in 
linear and nonlinear regimes, respectively. In Sec. V, a summary is presented.

%========================================================================
\section{Formalism}
%========================================================================
\subsection{Electron dynamics calculation in crystalline solids}
%Now correcting 2015/04/10

First we briefly explain a general feature of electron dynamics calculations 
in crystalline solids based on the TDDFT.
Details of the method are explained in Ref. \cite{ba00}.
We consider electron dynamics in crystalline solids under visible or infrared laser pulses. 
A typical wavelength of the electric field of such laser pulses is a micrometer, 
while the length scale of the electron dynamics induced by the laser pulse is 
less than a nanometer. Therefore, in a unit cell of crystalline solids, 
we can treat the laser electric field as a spatially-uniform  and time-dependent 
electric field $\vec E(t)$. This is a dipole approximation for interactions 
between the laser pulse and electrons in a medium. The electron dynamics is 
described by the following time-dependent Kohn-Sham (TDKS) equation,
\be
i\hbar \frac{\partial}{\partial t} \psi_i(\vec r,t)
= h_{KS}(t)\psi_i(\vec r,t),
\label{eq:tdks}
\ee
where $h_{KS}(t)$ is the time-dependent Kohn-Sham Hamiltonian.
It is given by
\be
h_{KS}(t)=\frac{\left \{ -i\hbar \vec \nabla +\frac{e}{c}\vec A(t)\right \}^2}{2m_e}
+\hat v_{ion}[\vec A(t)] + v_H(\vec r,t) + v_{xc}(\vec r,t),
\label{eq:tdks-h}
\ee 
where $\vec A(t)$ is the vector potential which is related to the electric field
$\vec E(t)$ by $\vec A(t) = -c\int^tdt' \vec E(t')$. 
The Hamiltonian contains the Hartree potential $v_H(\vec r,t)$,
an exchange-correlation potential $v_{xc}(\vec r,t)$, 
and an ionic potential $\hat v_{ion}[\vec A(t)]$.
The Hamiltonian of Eq. (\ref{eq:tdks-h}) is periodic in space so that
we may apply the Bloch theorem to the orbitals $\psi_i(\vec r,t)$ at
each time.
For the ionic potential, we employ a norm-conserving 
pseudopotential \cite{tr91} with the separable approximation \cite{kl82}. 
The ionic potential depends on the vector potential $\vec A(t)$
due to the nonlocality of the potential \cite{ba00}. 

The electric field  $\vec E(t)$ may include 
an induced polarization field which depends on macroscopic geometry
of the material which we treat. In the present calculation, we assume
the transverse geometry explained in Ref. \cite{ya12} in which the induced 
polarization fields do not appear.
Moreover, the vector potential $\vec A(t)$ may include exchange-correlation
terms in the time-dependent current density functional theory \cite{vi96,be07}. 
We simply ignore them, since reliable potentials have not yet 
been available at present.

We define a matter current density of electrons $\vec j (\vec r, t)$, 
\be
\vec j(\vec r, t)=\frac{1}{m_e}\sum_i \Re \left [
\psi^*_i(\vec r, t)\{-i\hbar \vec \nabla +\frac{e}{c}\vec A(t)\}\psi_i(\vec r, t)
\right].
\label{eq:current_dns}
\ee
and spatially averaged electric current density $\vec J(t)$., 
\be
\vec J(t) = -e \frac{1}{\Omega} \int_{\Omega} 
\vec j(\vec r,t) d\vec r 
+\vec J_{NL}(t),
\label{eq:current_elec}
\ee
where $\Omega$ is a volume of the unit cell and $\vec J_{NL}(t)$ is 
a contribution from the nonlocal part of the ionic pseudopotential 
\cite{ba00}.

%========================================================================
\subsection{Exchange-correlation potentials}
%========================================================================

We will use the LDA, TB-mBJ, and HSE potentials in the real-time TDDFT 
calculations in the adiabatic approximation. 
We present below brief explanations for the TB-mBJ and HSE 
potentials.

The TB-mBJ exchange potential in the adiabatic approximation 
has the following form,
\be
v^{ex}_{TB-mBJ}(\vec r, t)=c_m v_{BR}(\vec r, t)+(3c_m-2)\frac{1}{\pi}
\sqrt{\frac{5}{12}}\sqrt{\frac{\tau(\vec r,t)}{\rho(\vec r,t)}},
\label{eq:tm-mbj1}
\ee
where $v_{BR}(\vec r, t)$ is the Becke-Roussel potential \cite{be89},
$\rho(\vec r,t)=\sum_i |\psi_i(\vec r,t)|^2$ is the electron density, and 
$\tau(\vec r, t)$ is the kinetic energy density given by
\be
\tau(\vec r, t)= \frac{1}{m_e} \sum_i \left | 
\{-i \hbar \vec \nabla + \frac{e}{c}\vec A(t)\}\psi_i(\vec r, t)
\right|^2.
\label{eq:tm-mbj}
\ee

It has been known that the band gap depends crucially on the 
parameter $c_m$ of Eq. (\ref{eq:tm-mbj1}) \cite{ko12}. 
In the original TB-mBJ paper \cite{tr09}, a formula for the mixing 
parameter $c_m$ is proposed using the electron density and its gradient.
In this work, we treat the parameter $c_m$ as adjustable and determine the value
in such a way that the band gap coincides with the experimental gap of silicon.
As will be shown, we may find the value of $c_m$ that reproduces the experimental
optical gap as well as the band gap simultaneously.
The potential of Eq. (\ref{eq:tm-mbj1}) is not 
gauge invariant, since the kinetic energy density of Eq. (\ref{eq:tm-mbj}) is not 
\cite{ta05-1,ta05-2,pi09,ra10}.
To recover the gauge invariance, we replace all kinetic 
energy density terms in the potential as 
\be
\tau(\vec r, t) \rightarrow 
\tau(\vec r,t)-\vec j^2(\vec r, t)/\rho(\vec r,t),
\ee
where $\vec j(\vec r, t)$ is the current density of 
Eq. (\ref{eq:current_dns}).
For the correlation potential, we use the PW91 correlation 
potential \cite{pe92} following the original TB-mBJ paper \cite{tr09}.

The HSE exchange-correlation potential is given by 
\be
v_{xc} = v^{PBE}_{xc} - \frac{1}{4} v^{SR,PBE}_x + \frac{1}{4} 
\hat{v}^{SR,Fock},
\ee
where $v^{PBE}_{xc}$ is the PBE exchange-correlation potential 
\cite{pe96}, $v^{SR,PBE}_{x}$ and $v^{SR,Fock}$ are the PBE exchange 
and the non-local Fock-like potentials corresponding to the short-range part 
of the Coulomb potential. The HSE exchange-correlation potential includes 
a parameter $\mu$, the range separation parameter for the Coulomb 
interaction. In this work, we use the value $\mu$ of 0.3 ${\rm \AA ^{-1}}$.

%========================================================================
\subsection{Numerical method}
%========================================================================

We solve the TDKS equation, Eq. (\ref{eq:tdks}), in real time and 
real space \cite{va99,ya96}.
We choose a cubic unit cell containing eight 
silicon atoms and 32 valence electrons in total. 
We consider electron dynamics only, freezing atomic positions at 
their equilibrium positions in the ground state.
The unit cell is discretized by $16^3$ uniform grid points.
The first Brillouin zone is discretized by $24^3$ $k$-points for calculations
using the LDA and TB-mBJ potentials, and $12^3$ $k$-points when using
the HSE potential.
We use the preudopotential which was constructed using the LDA.

%========================================================================
\subsubsection{Predictor-corrector procedure}
%========================================================================

For time-evolution calculations, we repeat a time evolution of a small time step $\Delta t$:
\be
\psi_i(\vec r,t+\Delta t) \approx 
\exp\left[ \frac{h_{KS}(t+\frac{\Delta t}{2})}{i\hbar}\Delta t \right]\psi_i(\vec r,t),
\ee
where the Hamiltonian is approximately evaluated at a time $t+\Delta t/2$.
We expand the time evolution operator, 
$\exp\left[ h_{KS}(t+\frac{\Delta t}{2})\Delta t/i\hbar \right]$, in Taylor series and take
up to fourth order \cite{fl78,ya96}:
\be
\exp\left[ \frac{h_{KS}(t+\frac{\Delta t}{2})}{i\hbar}\Delta t \right]
\approx \sum^4_{n=0}\frac{1}{n!} 
\left(\frac{ h_{KS}(t+\frac{\Delta t}{2})}{i\hbar}\Delta t \right)^n.
\ee

In the above procedure, we need to guess the Hamiltonian at a time
$t+\Delta t/2$. In calculations using the LDA and HSE potentials,
we find that use of the Hamiltonian at time
$t$, $h_{KS}(t+\Delta t/2) \sim h_{KS}(t)$, provides results with sufficient
accuracy. However, we have found that
the predictor-corrector procedure to approximately evaluate 
$h_{KS}(t+\Delta t/2)$ is indispensable for stable time-evolution
when we employ the TB-mBJ potential. 
The predictor-corrector procedure is usually carried out in 
the following steps \cite{wa08}. 
We first calculate wave functions
$\psi^{pred}_i(\vec r, t+\Delta t)$ using the Hamiltonian at time $t$,
\be
\psi^{pred}_i(\vec r, t+\Delta t)=\exp \left[
\frac{h_{KS}(t) \Delta t}{i\hbar}
\right ] \psi_i(\vec r,t).
\ee
Using the Kohn-Sham orbitals $\psi_i^{pred}$, we next construct 
the Hartree potential $v^{pred}_H$ and the exchange 
correlation potential $v^{pred}_{xc}$ at time $t+\Delta t$.
We then approximate the Hamiltonian at time $t+\Delta t/2$ 
by averaging the potentials at time $t$ and $t+\Delta t$ 
as follows:
\be
v_H(\vec r,t+\frac{\Delta t}{2})\approx 
\frac{v^{pred}_H(\vec r,t+\Delta t)+v_H(\vec r,t)}{2},
\ee
\be
v_{xc}(\vec r,t+\frac{\Delta t}{2})\approx 
\frac{v^{pred}_{xc}(\vec r,t+\Delta t)+v_{xc}(\vec r,t)}{2}.
\ee
Finally, we calculate the Kohn-Sham orbitals at $t+\Delta t$ by Eq. (5) 
using the Hamiltonian at time $t+\Delta t/2$.

\begin{figure}[tb]
 \begin{center}
  \includegraphics[width= 6.5cm]{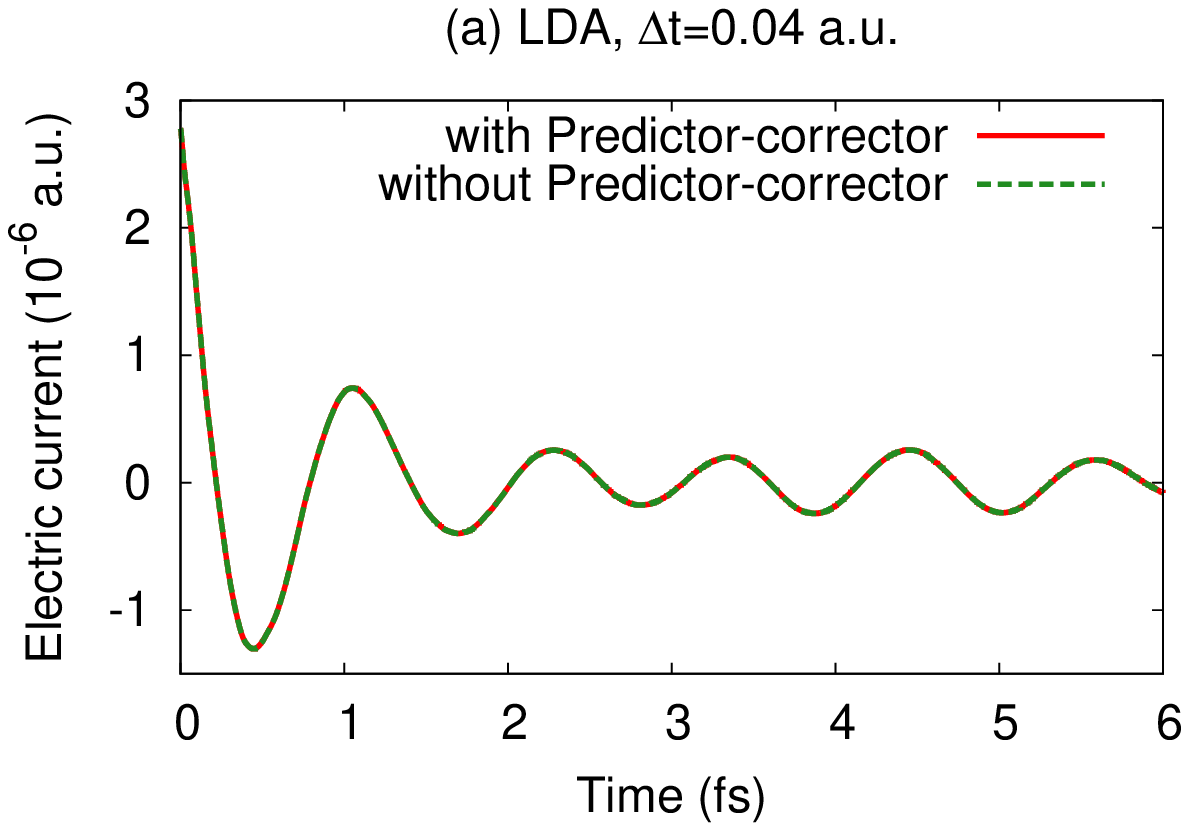}
  \includegraphics[width= 6.5cm]{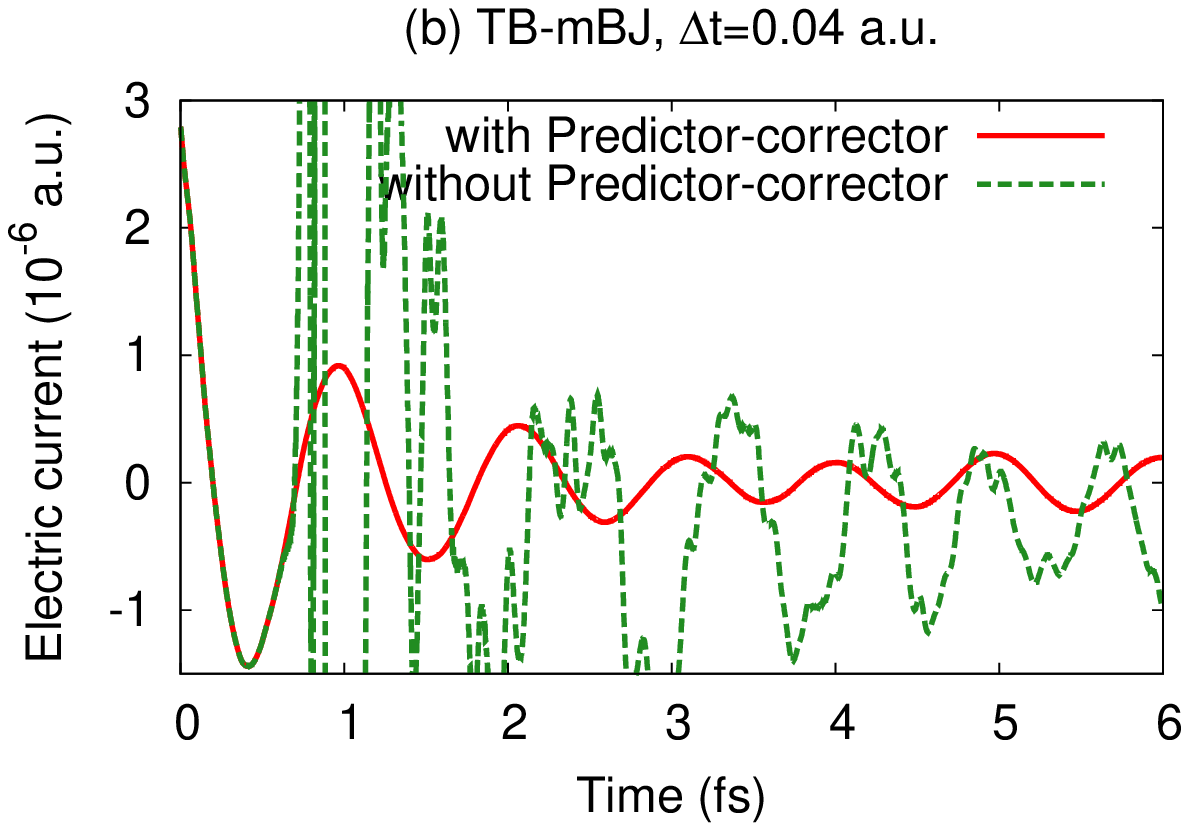}
 \end{center}

 \caption{(color online)
Electric currents as functions of time after an instantaneous weak 
distortion is applied at $t=0$. Panel (a) shows the results using 
the LDA potential, and panel (b) shows the results using 
the TB-mBJ potential. In both panels, calculated results with (red-solid)
and without (green-dashed) the predictor-corrector procedure are compared.}
 \label{fig:lin_current_mGGA_no_pred}
\end{figure}

In the following, we numerically demonstrate the necessity of the predictor-corrector 
procedure when
the TB-mBJ potential is used.
Figure \ref{fig:lin_current_mGGA_no_pred} shows the electric current 
in the crystalline silicon as a function of time after an instantaneous 
weak distortion is applied at $t=0$.  A time step of $\Delta t =0.04$ a.u. is used.
Figure \ref{fig:lin_current_mGGA_no_pred} (a) shows the electric current using the 
LDA potential, and (b) shows the electric current 
using the TB-mBJ potential. In both panels, red-solid lines show 
results using the predictor-corrector procedure, while green-dashed lines show 
results without using the predictor-corrector procedure. One sees that, 
while the calculations using the LDA potential give the same result 
regardless of whether the predictor-corrector procedure is used, 
the calculations using the TB-mBJ potential proceed stably only if 
the predictor-corrector step is used. Until $t=0.8$ fs 
(about 800 time steps),
the calculated electric currents with and without the predictor-corrector 
procedure coincide with each other.
However, at $t=0.8$ fs, the calculation without the predictor-corrector 
procedure starts to show unphysical oscillations.
We carried out calculations with and without the predictor-corrector
procedure for several time steps, $\Delta t=0.02, 0.04$, and $0.08$ a.u., and
have found that results without the predictor-corrector procedure
always fail showing unphysical oscillations after a certain number of iterations. 
Calculations using the predictor-corrector
procedure give the same result for the three time steps.
We thus conclude that the predictor-corrector procedure is indispensable 
for stable time-evolution of orbitals when using the TB-mBJ potential. 
In the calculations presented below, we use the predictor-corrector 
procedure when we employ the TB-mBJ potential.

\subsubsection{Nonlocal Fock-like term}
In calculations using the HSE potential, we need to calculate operations
of nonlocal Fock-like term on orbitals. The nonlocal Fock-like term 
$\hat{v}^{SR,Fcok}$ has the following form,
\be
\hat{v}^{SR,Fock} \psi_i(\vec r,t) = 
\sum_{j= occ} \int d\vec r' v_{SR}(\vec r - \vec r')
\psi_j(\vec r,t)\psi^*_j(\vec r',t) \psi_i(\vec r',t),
\ee
where only occupied orbitals are summed.
$v_{SR}(\vec r)$ is the short-ranged Coulomb-like potential, 
$v_{SR}(\vec r ) = {\rm erf}(\mu r)/r$.
In practical calculations, we treat the nonlocal Fock-like term 
in Fourier space as:
\be
\hat{v}^{SR,Fock} \psi_i(\vec r,t) =
\sum_{j= occ} \psi_j(\vec r,t) \frac{1}{(2\pi)^3}
\int d\vec G e^{-i\vec G \cdot \vec r} \tilde{v}^{SR}(\vec G) 
\tilde{\rho}_{ji}(\vec G,t),
\label{eq:Fock}
\ee
where $\tilde{v}^{SR}(\vec G)$ and $\tilde{\rho}_{ji}(\vec G,t)$ 
are Fourier transformation of $v^{SR}(\vec r)$ and orbital products 
$\rho_{ji}(\vec r,t) = \psi^*_j(\vec r,t)\psi_i(\vec r,t)$, 
respectively.

The Fourier space operation has been carried out efficiently 
employing the Fast-Fourier-Transformation library 
on large-scale accelerator type supercomputers.
In the HSE calculation, we further employ so-called
down-sampling method \cite{pa06} to reduce
the computational cost, which we briefly describe below.
First, we decompose the orbital index $i$ into the reciprocal lattice 
vectors of the unit cell $\vec k$ and  the band index $b$:
$\psi_{i}(\vec r,t) \rightarrow \psi_{\vec k b}(\vec r,t)$.
Using this notation, we rewrite Eq. (\ref{eq:Fock}) as follows:
\be
\hat{v}^{SR,Fock} \psi_{\vec k b}(\vec r,t) =
\sum_{\vec q, b'= occ} \psi_{\vec q b'}(\vec r,t) \frac{1}{(2\pi)^3}
\int d\vec G e^{-i\vec G \cdot \vec r} \tilde{v}^{SR}(\vec G) 
\tilde{\rho}_{\vec qb', \vec k b}(\vec G,t),
\label{eq:Fock-DS}
\ee
where the orbital product is described as 
$\rho_{\vec qb', \vec k b}(\vec r,t) = \psi^*_{\vec qb'}
(\vec r,t)\psi_{\vec k b}(\vec r,t)$ using the new indices.
In the down-sampling method, the summation of Eq. (\ref{eq:Fock-DS}) 
is carried out reducing the number of $\vec q$-points.
In this work, we sample $12^3$ $k$-points to describe Kohn-Sham 
orbitals, while we sample $4^3$ $q$-points 
to evaluate the operation of  the non-local Fock-like term 
of Eq. (\ref{eq:Fock-DS}).

%========================================================================
\section{Linear response calculation in real time}
%========================================================================

In this and next sections, we apply the real-time TDDFT calculations 
for electron dynamics in crystalline silicon using the LDA, TB-mBJ, and 
HSE potentials. 
Although linear response properties of solids may be investigated either in 
frequency or in time domains, time-domain calculation is a unique option for strongly 
nonlinear dynamics induced by ultra-short, highly-intense laser pulses.
In this section, we discuss a real-time calculation for a dielectric
function of silicon. The purpose of this section is to demonstrate that numerical methods 
of time evolution calculations presented in the previous section work stably 
in real-time linear response calculations.

For dielectric functions, several {\it ab-initio} approaches have been developed.
Among them,  {\it GW} plus Bethe-Salpeter 
equation approach \cite{ro98,ro00}, which is based on 
many-body perturbation theory, has been successful to describe 
dielectric functions of various materials quantitatively.  
{\it Ab-initio} calculations based on TDDFT have also been presented \cite{on02}.
Recently, dielectric functoins in the TDDFT have been reported using
meta-GGA \cite{na11} and HSE \cite{pa08} functionals, which are successful in describing
band gap energies of various materials correctly.

We briefly explain how we calculate dielectric function as a function of
frequency in real-time TDDFT calculations. 
We calculate time evolution of electron orbitals under a weak 
electric field, 
$\vec E(t) = -\frac{1}{c}\frac{\partial}{\partial t}\vec A(t)$, 
solving Eq. (\ref{eq:tdks}). Using electric conductivity $\sigma_{ij}(t)$,
the spatially-averaged electric current $\vec J(t)$, which is given by 
Eqs. (\ref{eq:current_dns}) and (\ref{eq:current_elec}), is related to 
the electric field $\vec E(t)$ by
\be
J_i(t) = \sum_j \int dt' \sigma_{ij}(t-t') E_j(t'),
\ee
where $i$ and $j$ indicate Cartesian component, $i,j=x,y,z$.
Taking Fourier transformations of both sides, we obtain a relation 
in frequency representation,
\be
J_i(\omega) = \sum_j \sigma_{ij}(\omega) E_j(\omega),
\label{eq:sigma}
\ee
where $E_i(\omega)$, $J_i(\omega)$, and $\sigma_{ij}(\omega)$ are 
Fourier transformations of $E_i(t)$, $J_i(t)$, and $\sigma_{ij}(t)$, respectively.
Frequency-dependent dielectric function $\epsilon_{ij}(\omega)$ is related
to $\sigma_{ij}(\omega)$ as usual,
\be
\epsilon_{ij}(\omega) = \delta_{ij}+\frac{4\pi i}{\omega} \sigma_{ij}(\omega).
\label{eq:epsilon}
\ee

In principle, we may use any electronic field $E_i(t)$ to calculate 
$\sigma_{ij}(\omega)$.
For numerical convenience, we employ a $\delta$-type distortion 
in $z$-direction for the electric field, $\vec E(t)=s \vec e_{z} \delta(t)$ where $s$ 
is a small parameter. Then the conductivity is proportional to the electric current, 
$\sigma_{zz}(t)=J_z(t)/s$. 
This relation tells us that the Fourier transformation of the electric 
current provides the frequency-dependent conductivity.
To reduce numerical noises caused by 
the finite evolution time $T$, 
we use a mask function $M(x)$ in the Fourier transformation as follows,
\be
\sigma_{zz}(\omega)=\frac{1}{s}  \int^{T}_0 dt  J_z(t) e^{i\omega t} M(t/T).
\label{eq:sigma2}
\ee
In practice, we employ the mask function $M(x)=1-3x^2+2x^3$ \cite{ya06}.

\begin{figure}[tb]
 \begin{center}
  \includegraphics[width= 6.5cm]{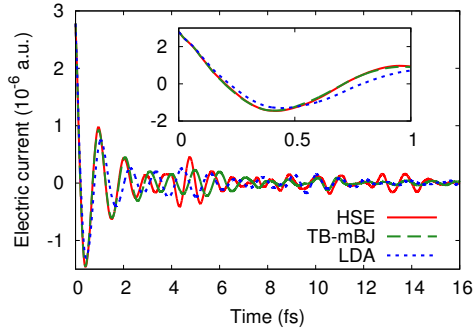}
 \end{center}

 \caption{(color online)
   Electric currents in crystalline silicon as functions of time after
   an impulsive distortion is applied at time $t=0$.
   Results using the three exchange-correlation potentials are compared:
   HSE (red-solid), TB-mBJ (green-dashed), and LDA (blue-dotted).}
 \label{fig:lin_current}
\end{figure}

Figure \ref{fig:lin_current} shows electric currents $J_z(t)$ in crystalline silicon 
as functions of time after the delta-function distortion is applied at $t=0$.
Red-solid line shows the result using the HSE potential, 
green-dashed line shows the result using the TB-mBJ potential, 
and blue-dotted line shows the result using the LDA potential. 
All currents accurately coincide with each other at $t=0$, 
since the initial current is determined by the sum rule \cite{ba00}. 
Shortly after the distortion (0 fs to 1 fs), 
all currents look similar to each other. 
They start to depart after a few femtoseconds. 
As discussed above, the electric current is proportional to 
the conductivity as a function of time when we employ the delta-type 
distortion. Therefore, the difference of the currents directly 
reflects the difference of the conductivities.

\begin{figure}[tb]
 \begin{center}
  \includegraphics[width= 6.5cm]{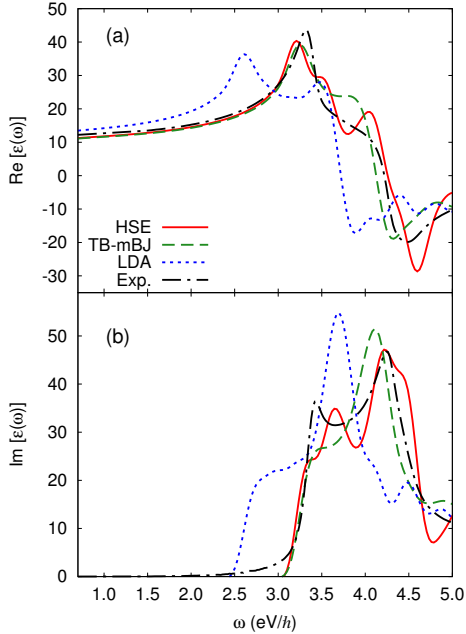}
 \end{center}

 \caption{Dielectric functions of silicon; real part in (a) and imaginary part in (b).
Results using the three exchange-correlation potentials are compared: 
HSE (red-solid), TB-mBJ (green-dashed), and LDA (blue-dotted).
An experimental result \cite{pa85} is also shown as black-dash-dotted lines.}
 \label{fig:eps_Re}
\end{figure}

From the currents shown in Fig. \ref{fig:lin_current}, dielectric functions can be calculated
using Eqs. (\ref{eq:epsilon}) and (\ref{eq:sigma2}). 
Figure \ref{fig:eps_Re} shows the results:
the real part in (a) and the imaginary part in (b).
Red-solid line shows the result using the HSE potential, 
green-dashed line shows the result using the TB-mBJ potential, 
and blue-dotted line shows the result using the LDA potential. 
An experimental result \cite{pa85} is also shown as black dash-dotted line.
Our results using the HSE and TB-mBJ potentials
shown in Fig. \ref{fig:eps_Re} almost coincide with previous calculations 
using frequency-domain methods \cite{pa08,na11}.

As seen from the panel (b), the optical gap is underestimated when the LDA
potential is used. This is closely related to the well-known band gap problem of the LDA. 
In contrast, the optical gap is well reproduced when we use the HSE
and TB-mBJ potentials. The real parts of the dielectric functions 
calculated using the HSE and TB-mBJ potentials also reproduce accurately 
the experimental values in low frequency region (below 3 eV). 

%========================================================================
\section{Nonlinear electron dynamics under intense laser pulses}
%========================================================================

In this section, we investigate nonlinear electron dynamics 
in crystalline silicon under intense, ultra-short laser 
pulses using the three exchange-correlation potentials.
We employ electric fields derived from the following vector potential,
\begin{eqnarray}
  A(t) = \left\{ \begin{array}{ll}
    -\frac{cE_0}{\omega}\cos(\omega t)\sin^2(\frac{\pi}{T_d}t) & 
(0  < t < T_d ) \\
    0 & ({\rm otherwise}),
  \end{array} \right.
\end{eqnarray}
where $\omega$ is a mean frequency of the field, $T_d$ is 
the pulse duration, and $E_0$ is the maximum electric field
of the pulse. We set $\omega$ and $T_d$ to $1.35$ eV/$\hbar$ and 
$16$ fs, respectively. We define the maximum intensity of the
laser pulse $I_0$ as $I_0=cE^2_0/8\pi$. 
We calculate electron dynamics using three exchange-correlation 
potentials changing the intensity $I_0$.
We will examine two physical quantities: the excitation energy during and after 
the laser irradiation, and the number density of excited electrons after 
the laser irradiation.

%========================================================================
\subsection{Excitation energy}
%========================================================================

We first consider electronic excitation energy induced by laser irradiation.
The excitation energy is one of the most important quantities to 
investigate laser-matter interactions. For example, optical damage 
thresholds can be estimated by comparing the electronic excitation 
energy with cohesive energy of solids \cite{le14,sa15-eprint}. 
Although the LDA and HSE potentials are explicitly derived from
energy functionals, the TB-mBJ potential is not derived from any functionals
but given directly. It is even unclear whether there exists 
an energy functional which provides the TB-mBJ potential. We first 
discuss how to evaluate electronic excitation energy when we use 
the TB-mBJ potential. We then investigate differences in electronic 
excitation energies using the three exchange-correlation potentials.

We first note that there are two procedures to calculate electronic excitation 
energy which should give the same results if the exchange-correlation 
potential is derived from an energy functional. One is to use the explicit 
expression of the energy functional given by
\begin{eqnarray}
E_{ex}(t)&=&\sum_i \int_{\Omega} d\vec r \psi^*_i(\vec r,t)
\left[
\frac{\left ( -i\hbar \nabla +\frac{e}{c}\vec A(t)\right )^2}{2m}
+\hat v_{ion}[\vec A(t)]
\right]\psi_i(\vec r,t) \nonumber \\
&+&\frac{1}{2}\int d\vec r v_H(\vec r, t)\rho(\vec r,t)
+E_{xc}[\{ \psi_i(\vec r,t) \}] - E_{GS},
\label{eq:Eex01}
\end{eqnarray}
where $E_{xc}$ is the exchange-correlation energy functional and 
$E_{GS}$ is the energy of the ground state.
The other is to calculate the work done on electrons by the laser electric
field,
\be
W(t) =  \int^t_{-\infty} dt' \vec J(t') \cdot \vec E(t'),
\label{eq:Eex02}
\ee
where $\vec J(t)$ is the electric current defined 
by Eq. (\ref{eq:current_elec}).
When the exchange-correlation potential is derived from the energy
functional $E_{ex}[\{\psi_i(\vec r,t)\}]$, 
we may easily prove that there holds
\be
W(t)=\int^t_{-\infty} dt' \frac{d}{dt'}E_{ex}(t') =E_{ex}(t).
\ee
In calculations using the LDA and HSE potentials, we may employ
either Eq. (\ref{eq:Eex01}) or Eq. (\ref{eq:Eex02}). 
In calculations using the TB-mBJ potential, Eq. (\ref{eq:Eex02})
is the unique option.

\begin{figure}[tb]
 \begin{center}
  \includegraphics[width= 6.5cm]{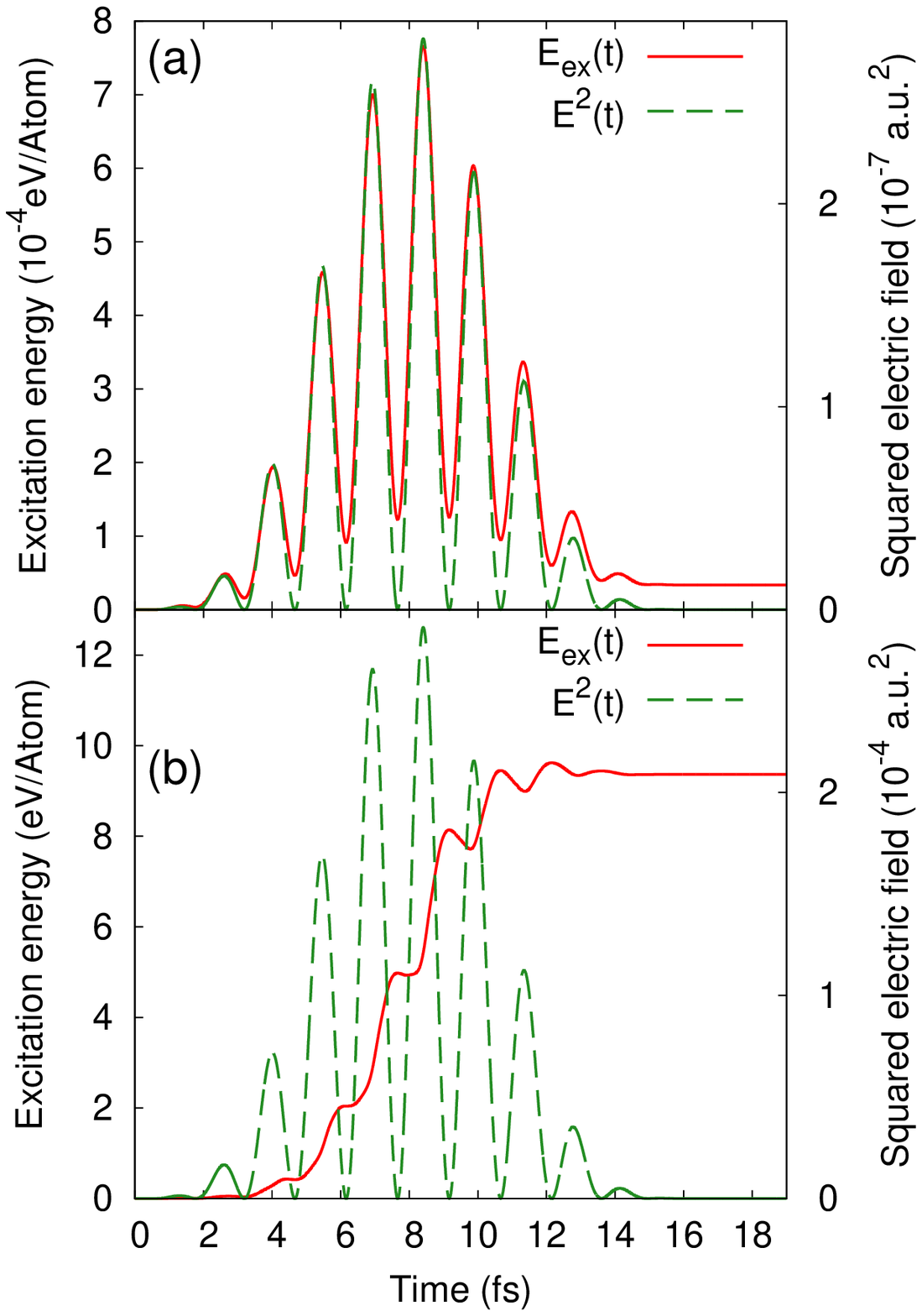}
  \includegraphics[width= 6.5cm]{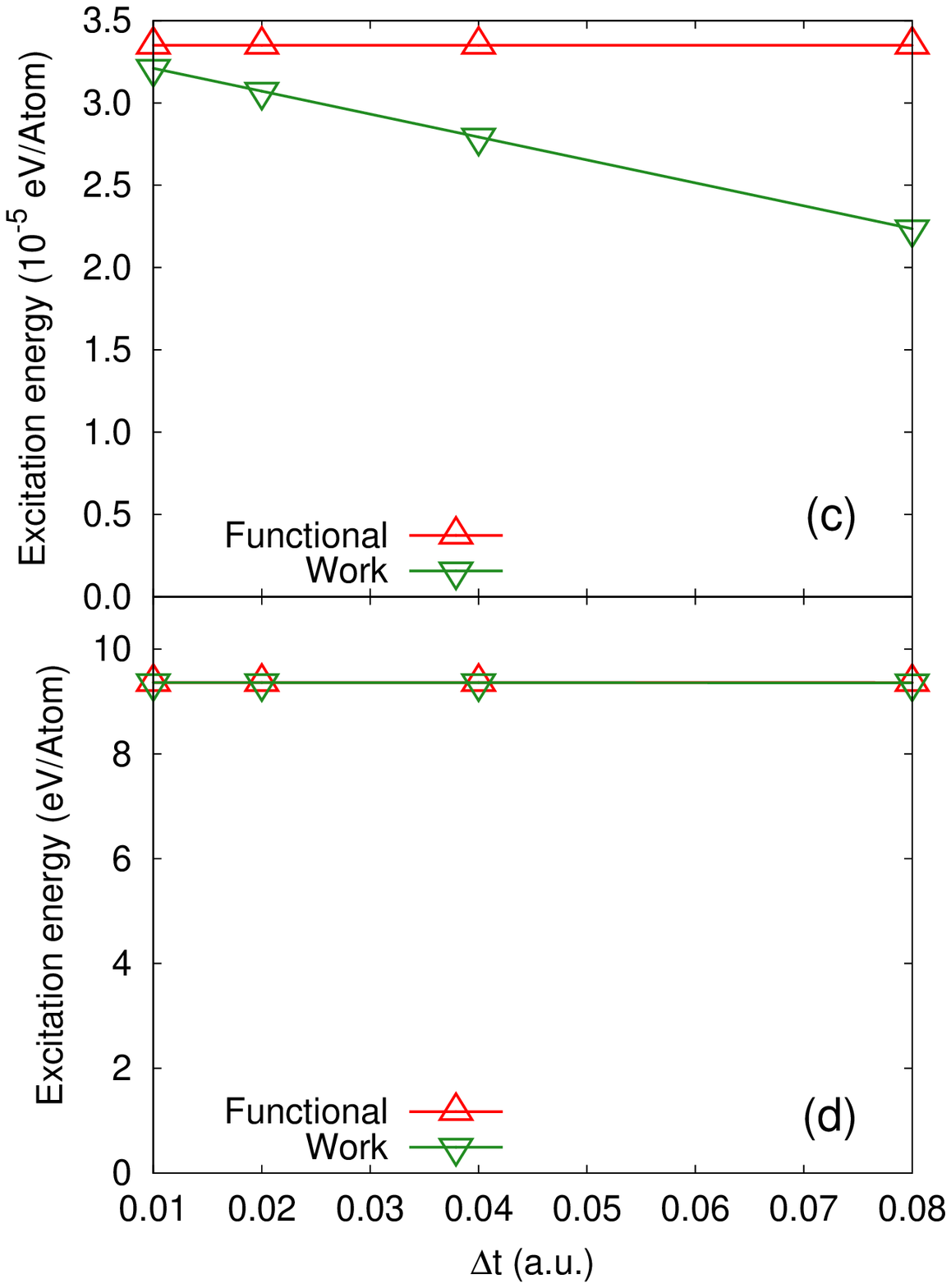}
 \end{center}

 \caption{
Laser-induced electronic excitation energy calculated using 
the LDA potential. Panels (a) and (b) show the electronic 
excitation energy (red-solid) and the applied squared electric field (green-dashed)
as functions of time. Panels (c) and (d) show the electronic 
excitation energy after the laser irradiation ends
using Eqs. (\ref{eq:Eex01}) and (\ref{eq:Eex02}).
Results in the panels (a) and (c) are for the irradiation of the weak laser pulse
($I_0=1.0\times10^{10}$W/cm$^2$), while results in the panels (b) and (d)
are for the irradiation of the strong laser pulse
($I_0=1.0\times10^{13}$W/cm$^2$).
}
 \label{fig:dtconv_p1d13_BJ}
\end{figure}

Although Eq. (\ref{eq:Eex01}) and Eq. (\ref{eq:Eex02}) are analytically
equivalent to each other if the potential is derived from an energy functional,
we have found that Eq. (\ref{eq:Eex01}) is numerically more favorable
when the applied laser pulse is weak.
Figure \ref{fig:dtconv_p1d13_BJ} shows $E_{ex}(t)$ of Eq. (\ref{eq:Eex01}) 
using the LDA potential as a function of time for two intensities, 
$I=1.0 \times 10^{10}$ W/cm$^2$ in (a) and $I=1.0 \times 10^{13}$ W/cm$^2$ 
in (b). In the panel (a), we find an appreciable excitation energy during 
irradiation of the laser pulse. However, at the end of the pulse, 
the system almost returns to the ground state since the laser frequency is lower 
than the direct band gap. We shall call it {\it virtual} excitation. 
If we calculate the excitation energy using 
Eq. (\ref{eq:Eex02}), the integrand $\vec J(t) \cdot \vec E(t)$ behaves oscillatory 
during the irradiation of the laser pulse. The integration over time mostly cancels 
and gives the total work done by the laser pulse which is very small 
as seen from the panel (a).
Therefore, the calculation using Eq. (\ref{eq:Eex02}) may suffer from 
an accumulation of a numerical error during the time integral. 
Figure \ref{fig:dtconv_p1d13_BJ} (c) compares the electronic excitation 
energies after the weak laser pulse ends ($I_0=1.0\times 10^{10}$W/cm$^2$) 
using several different time steps $\Delta t$.
Red up-pointing and green down-pointing triangles show the excitation energies
calculated by Eq. (\ref{eq:Eex01}) and by Eq. (\ref{eq:Eex02}), respectively.
As seen from the figure, results coincide 
numerically if we employ sufficiently small time step. The error in 
the excitation energy increases linearly with the time step $\Delta t$
when we employ Eq. (\ref{eq:Eex02}). 

When the laser pulse is so strong that electrons are substantially excited
by the laser pulse, we find a steady increase of the excitation energy in time, as
seen in Fig. \ref{fig:dtconv_p1d13_BJ} (b). 
We shall call it {\it real} excitation. In this case, the difference in 
the numerical calculation is very small between the calculations using 
Eqs. (\ref{eq:Eex01}) and (\ref{eq:Eex02}), 
even if we employ the largest time step, $\Delta t=0.08$ a.u.

When we use the TB-mBJ potential, we can only evaluate excitation energy
using Eq. (\ref{eq:Eex02}). To obtain excitation energy, 
we carry out two calculations using different time steps, $\Delta t = 0.02$ a.u. 
and $\Delta t = 0.01$ a.u. We then estimate the converged excitation 
energy by a linear extrapolation using the two calculations. 

\begin{figure}[tb]
 \begin{center}
  \includegraphics[width= 8cm]{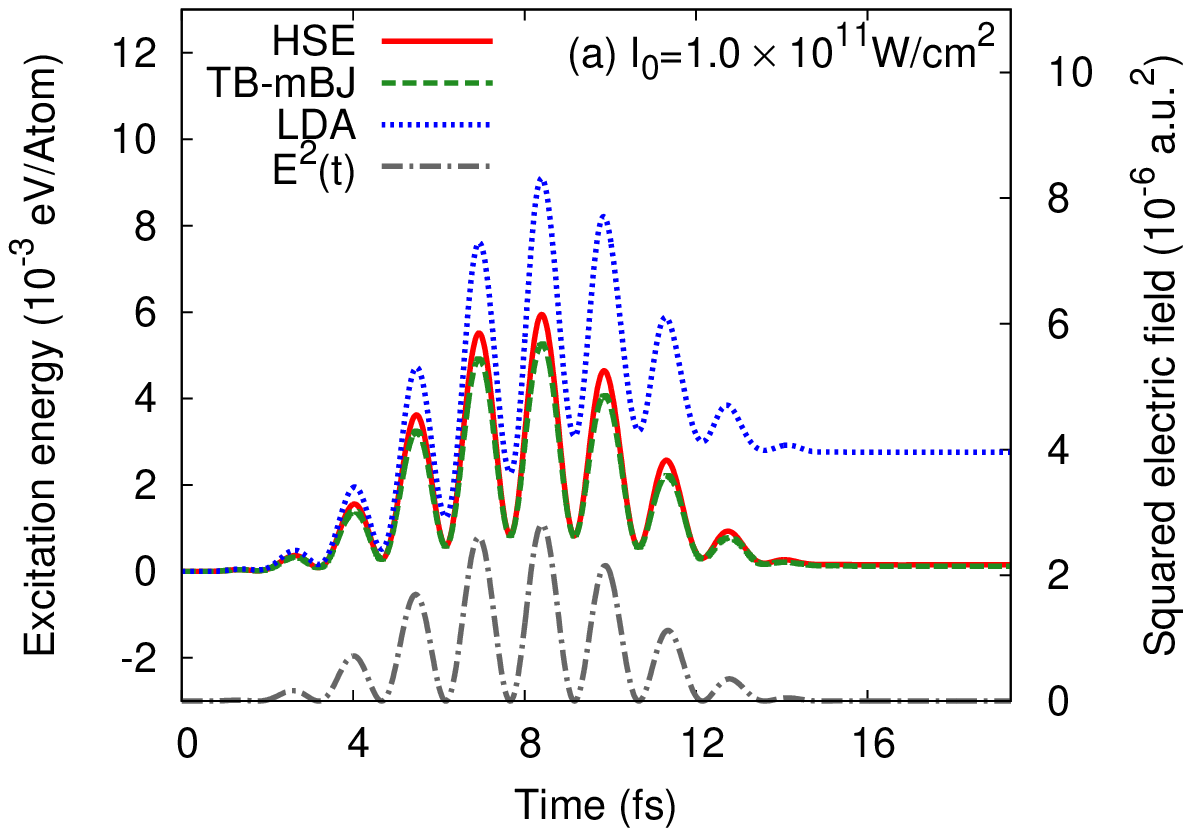}
  \includegraphics[width= 8cm]{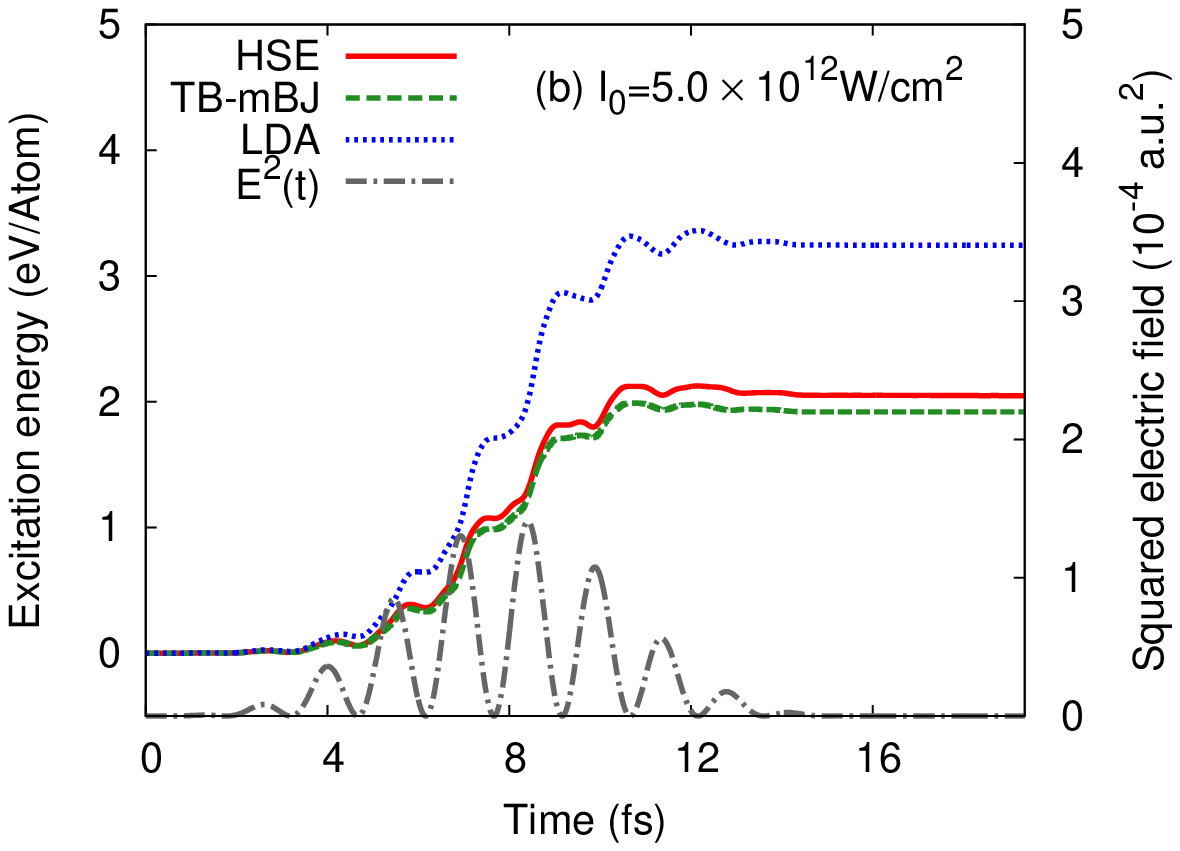}
 \end{center}
 \caption{(color online)
Electronic excitation energies as functions of time 
induced by laser irradiation. Panel (a) shows the excitation
energy under irradiation of a weak laser pulse 
($I_0=1.0\times 10^{11}$ W/cm$^2$), while panel 
(b) shows that of a strong laser pulse
($I_0=5.0 \times 10^{12}$ W/cm$^2$).
Results using the three exchange-correlation potentials  are compared: 
HSE (red-solid), TB-mBJ (green-dashed), and LDA (blue-dotted).
Squared electric fields are also shown as gray-dash-dotted lines.
}
 \label{fig:Si_Eex_t_p1d11}
\end{figure}

Figure \ref{fig:Si_Eex_t_p1d11} shows electronic excitation energies as 
functions of time using the three exchange-correlation potentials; 
LDA, TB-mBJ, and HSE. Figure \ref{fig:Si_Eex_t_p1d11} (a) shows 
results for a weak laser pulse ($I_0=1.0\times 10^{11}$W/cm$^2$), 
and Figure \ref{fig:Si_Eex_t_p1d11} (b) shows results for a strong laser pulse 
($I_0=5.0\times 10^{12}$W/cm$^2$). Squared electric fields $E^2(t)$ 
are also shown as gray dash-dot lines in each panel. 
As seen from Fig. \ref{fig:Si_Eex_t_p1d11} (a), the electric field
induces virtual excitation during the laser irradiation. 
After the irradiation ends, the excitation energies using
the HSE and TB-mBJ potentials return to almost zero. 
In contrast, the excitation energy using the LDA potential 
remains finite after the irradiation, showing a substantial real excitation. 
These results indicate that wider optical gap in the calculations using 
the TB-mBJ and HSE potentials suppresses
the real excitation (see Fig. \ref{fig:eps_Re}). 

Figure \ref{fig:Si_Eex_t_p1d11} (b) shows that electronic excitations 
induced by the strong electric field persist even after the pulse ends
irrespective of the exchange-correlation potentials.
One sees that the excitation energies using 
the HSE and TB-mBJ potentials are similar to each other,
while the excitation energy using the LDA potential is 
larger than the others at each time. 
This is consistent with the above finding that the wider band gap 
suppresses the real excitation. 

\begin{figure}[tb]
 \begin{center}
  \includegraphics[width= 6.5cm]{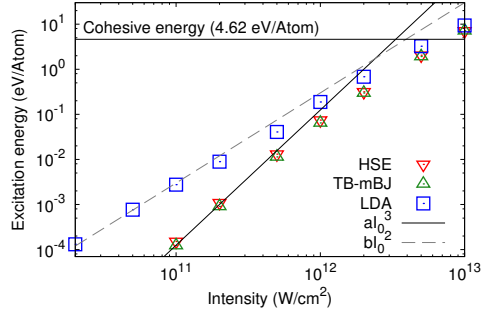}
 \end{center}

 \caption{(color online)
Excitation energies as functions of laser intensity
calculated using the three exchange-correlation potentials:
HSE (red down-pointing triangle), TB-mBJ (green up-pointing triangle), and 
LDA (blue square). Cubed and squared intensities normalized to the excitation energies
at low intensity are also shown as black-solid and gray-dashed lines, respectively.}
 \label{fig:Si_Eex}
\end{figure}

Next we investigate the electronic excitation energy after laser pulses end,
changing the laser intensity $I_0$. Figure \ref{fig:Si_Eex} shows excitation 
energies after the pulses end 
($t = 19.3$ fs).
Red down-pointing triangle shows the HSE result, green up-pointing 
triangle shows the TB-mBJ result, and blue square shows the LDA
result. Lines of cubed and squared intensity of the laser pulses
are also shown as black-solid and gray-dashed lines, respectively.
These lines are normalized at low intensity to coincide with 
the excitation energies.
Black horizontal line shows the cohesive energy of silicon,
$4.62$eV/atom \cite{fa91}, which may be regarded as a reference
of the damage threshold.

%As seen from Fig. \ref{fig:Si_Eex}, the excitation energy
%induced by the laser pulses is monotonically increases with 
%increasing the laser intensity $I_0$ in all cases.

Multi-photon excitations are expected 
for weaker intensities, while tunneling excitations are expected 
for stronger intensities \cite{ke64}.
At low intensity region, the excitation energy using 
the LDA potential can be fit with the squared intensity line $I^2_0$, 
while the excitation energies calculated using 
the TB-mBJ and HSE potentials can be fit with the cubed intensity line $I^3_0$.
These behaviors indicate that two and three-photon absorption processes
take place at the low intensity region for the LDA and the others cases, 
respectively. The numbers of absorbed photons are consistent with
the ratio of the photon energy of the laser pulse ($\hbar \omega =$1.35 eV) 
to the optical gap: the optical gap of silicon is 2.5 eV 
in the LDA case, and it is 3.1 eV in the HSE and 
the TB-mBJ cases (see Fig. \ref{fig:eps_Re}). 

One sees that the excitation energy using the LDA potential
is much higher than the excitation energies using 
the HSE and TB-mBJ potentials at low intensity region.
In contrast, the difference among the excitation energies becomes relatively small 
at high intensity region close to the cohesive energy of the medium.

%========================================================================
\subsection{Number density of excited electrons}
%========================================================================

The number density of excited electrons due to applied electric fields 
is one of significant observables to characterize laser excitation  processes
\cite{so00,te06}.
To define the number density of excited electrons after laser irradiation,
we first define eigenstates of the Kohn-Sham Hamiltonian after the 
laser pulse ends ($t_f =19.3$fs),
\be
h_{KS}(t_f)\phi_i(\vec r,t_f) = \epsilon^{t_f}_i \phi_i(\vec r,t_f).
\ee
Using these eigenstates, we define the number density of 
excited electrons $n_{ex}$ by,
\be
n_{ex} = n_{elec} - \frac{1}{\Omega}
\sum_{i,j = occupied}|\langle \phi_i(t_f)|
\psi_j(t_f)\rangle|^2,
\label{eq:nex_dns}
\ee
where $n_{elec}$ is the number density of valence electrons in the ground state, 
$\Omega$ is the volume of the unit cell, and 
$\psi_i(\vec r,t)$ are the time-dependent Kohn-Sham orbitals, 
which are solutions of Eq. (\ref{eq:tdks}).
Only occupied states are summed in Eq. (\ref{eq:nex_dns}).

\begin{figure}[tb]
 \begin{center}
  \includegraphics[width= 6.5cm]{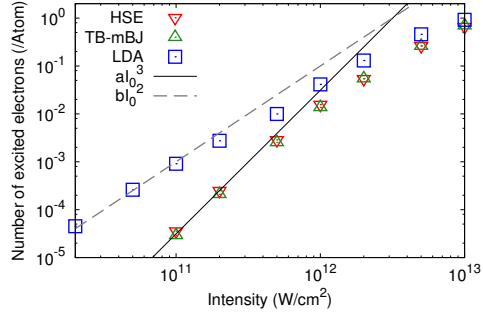}
 \end{center}

 \caption{(color online)
The number densities of excited electrons as 
functions of laser intensity calculated using 
the three exchange-correlation potentials:
HSE (red down-pointing triangle), TB-mBJ (green up-pointing triangle), and 
LDA (blue square). Cubed and squared intensities normalized to the number densities
of excited electrons at low intensity are also shown as black-solid and gray-dashed
lines, respectively.
}
 \label{fig:Si_Nex}
\end{figure}

Figure \ref{fig:Si_Nex} shows the number density of excited 
electrons as functions of laser intensity $I_0$.
We find similar features to those seen in excitation energies shown in 
Fig. \ref{fig:Si_Eex}.
The number density using the HSE potential is very close to 
the number density using the TB-mBJ potential. 
The number density using the LDA potential is larger
than the others.  While the number density using the LDA
potential is proportional to the squared intensity of laser pulse in
the low intensity region, the number densities using
the HSE and TB-mBJ potentials are proportional to the cubed intensity.
At high intensity region, the number densities of excited electrons 
using three exchange-correlation potential are similar to
each other.

\begin{figure}[tb]
 \begin{center}
  \includegraphics[width= 6.5cm]{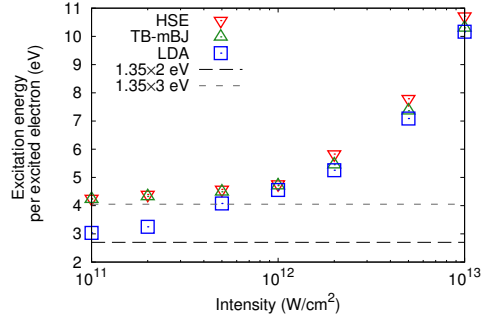}
 \end{center}

 \caption{(color online)
Excitation energy per excited electron
is plotted against the laser intensity calculated using 
the three exchange-correlation potentials:
HSE (red down-pointing triangle), TB-mBJ (green up-pointing triangle), and 
LDA (blue square). 
Two horizontal lines are shown at $2 \times 1.35$ (black-dashed)
and $3 \times 1.35 $ eV (gray-dotted).}
 \label{fig:Si_Nex_Eex}
\end{figure}

Figure \ref{fig:Si_Nex_Eex} shows excitation energy per excited 
electron, namely, the excitation energy shown in Fig. \ref{fig:Si_Eex}
divided by the number of excited electrons shown in Fig. \ref{fig:Si_Nex}
as a function of laser intensity $I_0$. Twice and three times of 
the photon energy, 1.35 eV, are also shown by horizontal lines. 
At low intensity region, the excitation energy per excited electron 
using the TB-mBJ and HSE potentials approach to three times of 
the photon energy, 
$3\times1.35$ eV, while that using the LDA potential approaches 
twice of the photon energy, 
$2\times1.35$ eV. This fact obviously indicates that two and three-photon
absorption processes dominate in the LDA case and
the other cases, respectively, and is consistent with the intensity dependences 
seen in Figs. \ref{fig:Si_Eex} and \ref{fig:Si_Nex}.

%========================================================================
\section{Summary}
%========================================================================
In this paper, we developed methods to carry out real-time TDDFT calculations 
using the TB-mBJ meta-GGA potential and the HSE hybrid functional, which 
are known to reasonably reproduce band gaps of insulators.
To carry out stable time evolution of orbitals using the TB-mBJ potential,
we found that the predictor-corrector procedure is indispensable.
Without the predictor-corrector procedure, 
the electric current induced by an impulsive excitation
shows unphysical oscillations after the time evolution of a few femtoseconds. 
This failure cannot be avoided even if we employ very small time step.
In the calculations using hybrid functional, we developed a Fourier-space method 
to operate Fock-like term of hybrid functional on Kohn-Sham orbitals, 
making use of recently developed accelerator-type supercomputers. 

We examined electron dynamics using the TB-mBJ, HSE, and LDA potentials.
In linear response calculations, we investigated the dielectric
response of crystalline silicon in time-domain. 
Calculated dielectric functions using the HSE and TB-mBJ 
potentials are confirmed to be similar to the previous results of frequency-domain 
TDDFT calculations 
\cite{pa08,na11}.
In electron dynamics calculations under strong, untra-short laser pulses, 
we investigated the excitation energy and the number density of excited electrons 
during and after the laser irradiations. Since an explicit form of 
the energy functional corresponding to the TB-mBJ potential is not known, 
we developed a method to evaluate electronic excitation energy, not
referring to the energy functional but calculating a work done 
by the external field to electrons.
For irradiation of weak laser pulses, results using the HSE 
and TB-mBJ potentials are almost the same to each other, 
while the excitation energy and the number density of excited electrons 
of the LDA calculation are much larger than the others.
For irradiation of strong laser pulses close to the damage threshold,
we found that results are similar among the three potentials.

We consider that real-time TDDFT calculations using sophisticated 
exchange-correlation potentials such as the TB-mBJ and HSE potentials, 
which describe the band gap of various insulators reasonably, 
will enables us to carry out quantitative descriptions for electron dynamics
in crystalline solids even under extremely nonlinear conditions. 
They are expected to provide significant insights and knowledge 
in wide fields of optical sciences including nonlinear 
optical responses, optical-control of electrons, and laser processing in solids.

%========================================================================
\section*{Acknowledgments}
%========================================================================
S.A.S thanks K. Sekizawa and G. Wachter for helpful discussion.
This work is supported by the Japan Society of the Promotion
of Science (JSPS) Grants-in-Aid for Scientific Research Grant
Nos. 15H03674 and 25800124, and by the JSPS Grants-in-Aid for JSPS Fellows Grant
No. 26-1511.
The numerical calculations were performed on the supercomputer 
at the Institute of Solid State Physics, University of Tokyo, 
T2K-Tsukuba and HA-PACS at the Center for Computational Sciences, 
University of Tsukuba.

%========================================================================
%========================================================================

\end{document}